\documentclass[preprint2]{aastex}
\usepackage{natbib}

\newcommand{\vs}           {{\it vs.}}
\newcommand{\AWP}     {WISEA}
\newcommand{\obj}{W0707+1705}
\newcommand{\asec}      {\mbox{$^{\prime\prime}$}}
\newcommand{\um}         {\mbox{$\mu$m}}

\newcommand{\be}           {\begin{equation}}
\newcommand{\ee}           {\end{equation}}
\newcommand{\bea}          {\begin{eqnarray}}
\newcommand{\eea}          {\end{eqnarray}}
\newcommand{\WISE}       {{\sl WISE}}

\slugcomment{submitted to the Astronomical Journal}

\shorttitle{High Proper Motion Star}
\shortauthors{Wright et al.}

\begin{document}

\title{The First AllWISE Proper Motion Discovery: \AWP\ J070720.50+170532.7}

\author{Edward L.\ Wright\altaffilmark{1},
J.\ Davy Kirkpatrick\altaffilmark{3},
Christopher R.\ Gelino\altaffilmark{3},
Sergio Fajardo-Acosta\altaffilmark{3},
Gregory Mace\altaffilmark{1},
Peter R.\ Eisenhardt\altaffilmark{4},
Daniel Stern\altaffilmark{4},
Ian S.\ McLean\altaffilmark{1},
M.\ F.\ Skrutskie\altaffilmark{2},
Apurva Oza\altaffilmark{2},
M.\ J.\ Nelson\altaffilmark{2},
Michael C.\ Cushing\altaffilmark{6},
I.\ Neil Reid\altaffilmark{7},
Michele Fumagalli\altaffilmark{8,9,10},
Adam J.\ Burgasser\altaffilmark{11}
\altaffiltext{1}{UCLA Astronomy, PO Box 951547, Los Angeles CA 90095-1547}
\altaffiltext{2}{Department of Astronomy, University of Virginia, Charlottesville, VA, 22904}
\altaffiltext{3}{Infrared Processing and Analysis Center,
California Institute of Technology, Pasadena CA 91125}
\altaffiltext{4}{Jet Propulsion Laboratory, California Institute of Technology, 4800 Oak Grove Dr., Pasadena, CA, 91109, USA}
\altaffiltext{5}{National Radio Astronomy Observatory, Charlottesville, VA 22903}
\altaffiltext{6}{Department of Physics and Astronomy, University of Toledo, 2801 W. Bancroft St., Toledo\ OH\ 43606-3328}
\altaffiltext{7}{STScI, Baltimore MD}
\altaffiltext{8}{Carnegie Observatories, 813 Santa Barbara Street, 
  Pasadena, CA 91101, USA}
  \altaffiltext{9}{Department of Astrophysics, Princeton University, Princeton, NJ 
08544-1001, USA}
\altaffiltext{10}{Hubble Fellow}
  \altaffiltext{11}{University of California San Diego, 9500 Gilman Drive, La Jolla, CA 92093}
}

\email{wright@astro.ucla.edu}

\begin{abstract}

While quality checking a new motion-aware co-addition of all 12.5 months
of WISE data, the source WISE J070720.48+170533.0 was found to have
moved 0.9\asec\ in 6 months.  Backtracking this motion allowed us to identify
this source as 2MASS J07071961+1705464 and with several entries in the
USNO B catalog.  
An astrometric fit to these archival data gives a proper motion of
$\mu = 1793\pm2$~mas/yr and a parallax of $\varpi = 35 \pm 42$~mas.
Photometry from WISE, 2MASS and the POSS can be fit
reasonably well by a blackbody with $T = 3658$~K and an angular radius of 
$4.36 \times 10^{-11}$ radians.  No clear evidence of H$_2$ collision-induced
absorption is seen in the near-IR.  An optical spectrum 
shows broad deep CaH bands at 638 \& 690 nm, broad deep Na D
at 598.2~nm, and weak or absent TiO, indicating that this
source is an ultra-subdwarf  M star with a radial velocity 
$v_{rad} \approx -21 \pm 18$~km/sec
relative to the Sun.  Given its apparent magnitude, the distance is about
$39 \pm 9$ pc and the tangential velocity
is probably $\approx 330$~km/sec, but a more precise parallax is needed
to be certain.

\end{abstract}

\keywords{brown dwarfs -- infrared:stars -- solar neighborhood -- stars:late-type -- stars:low-mass -- stars:individual}

\section{Introduction}

The Wide-field Infrared Survey Explorer (\WISE) \citep{wright/etal:2010}
mapped the entire sky between 14 Jan 2010 and 17 Jul 2010 in 4 thermal infrared
bands centered at 3.4, 4.6, 12 \& 22 \um, then
continued on to map the entire sky again prior to 1 Feb 2011.  On
7 Aug 2010 the outer cryogen tank ran out of solid hydrogen coolant,
and the 3-band cryo phase began, where the 22 \um\ channel was lost
and the 12 \um\ channel had much reduced sensitivity due to the
background radiated by the optics which warmed to 45 K.  On 30 Sep
2010 the inner cryogen tank ran out of solid hydrogen, and the two
band post-cryo phase began.  This phase lasted until 1 Feb 2011,
with only the 3.4 and 4.6 \um\  channels operating.  This last phase
of the mission, which lasted for 4 months, was never co-added or
catalogued.  Only single frame detections were found to search for
Near Earth Objects as part of the NEOWISE project\citep{mainzer/etal:2011a}.

The AllWISE project is an effort to co-add all the WISE frames,
find peaks in these coadds, and then fit for the properties of the
source or sources responsible for each peak by doing profile fit
photometry using the pixels on each frame that covers the position
of the peak.  While doing this fit, the AllWISE project is allowing
for a motion in addition to the usual parameters of the position
and the fluxes in each of the 4 bands.

Previous efforts to find nearby stars with the WISE data have
concentrated on color-selected searches.  \citet{kirkpatrick/etal:2011}
found over one hundred brown dwarfs using the color cut W1-W2 $>
2$, and many more have been studied by other papers such as \citet{mace/etal:2013}.
These color-selected searches have found many high proper motion
objects.  However, several papers have searched for moving objects by
using the 2MASS match (3\asec\ radius) in the WISE catalog as a
veto \citep{gizis/troup/burgasser:2011, liu/etal:2011, thompson/etal:2013}.
With AllWISE we seek to find previously missed high motion objects
using positive associations between observations over the full time
span of WISE.  This has been done for objects bright enough to
appear in WISE single frame detections by \citet{luhman:2013},
leading to the discovery of a brown dwarf binary only 2 pc from the
Sun.  AllWISE is doing a similar search using coadded frames to
reach the full sensitivity of WISE.  \citet{kirkpatrick/etal:2014}
will discuss a large sample of AllWISE motion discoveries, but this
brief note describes one of the most interesting AllWISE motion
selected source whose colors are similar to hundreds of millions
of WISE objects.

\section{Observations}

\subsection{Astrometry \& Photometry}

\begin{figure}[tb]
\plotone{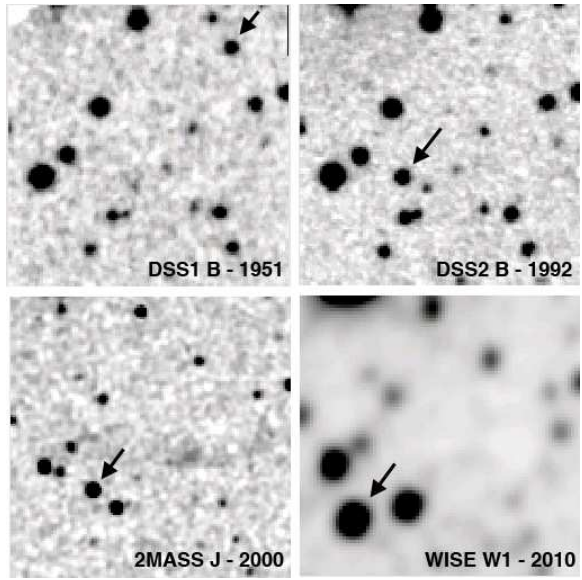}
\caption{$2 \times 2$ arcmin images of the field near WISE 0707+1705. 
North is up and East is to the left. Arrows point to the high proper motion 
star at each epoch.\label{fig:0707+1505-POSS12-W1}}
\end{figure}

\begin{table*}[tb]
\caption{Astrometric Data [J2000] \label{tab:astrometry}}
\begin{tabular}{rrrrrl}
MJD & $\alpha$ $[{}^\circ]$ & $\sigma_\alpha$ [mas] & $\delta$ $[{}^\circ]$ & $\sigma_\delta$ [mas] & Source \\
\hline
33957.456  & 106.814542  & 195 & 17.113712 & 79 & USNO B1.0 \\
48986.792  & 106.829188  & 333 & 17.098708 & 333 & POSS2 B FITS WCS \\
49277.000  & 106.829498  & 114 & 17.098514 & 25 & USNO B1.0 \\
50405.411  & 106.830604 & 333 & 17.097333 & 333 & POSS2 R FITS WCS \\
50923.000  & 106.831139 & 999 & 17.096873 & 999 & USNO B1.0 \\
51496.920  & 106.831671 & 333 & 17.096272 & 333 & POSS2 IR FITS WCS \\
51579.657  & 106.831722  & 70 & 17.096224 & 70 & 2MASS \\
51902.000  & 106.832112 &122 & 17.095854 & 122 & UCAC4 \\ 
52064.000  & 106.832222  & 88 & 17.095716 & 63 & CMC14 \\
55287.577  & 106.835361 & 90 & 17.092520 & 90 & WISE allsky \\
55479.133  & 106.835568 & 90 & 17.092344 & 90 & WISE postcryo \\
56545.350   & 106.836599 & 50 & 17.091278 & 50 & FanCam \\
\hline
\end{tabular}
\end{table*}

\begin{figure}[tb]
\plotone{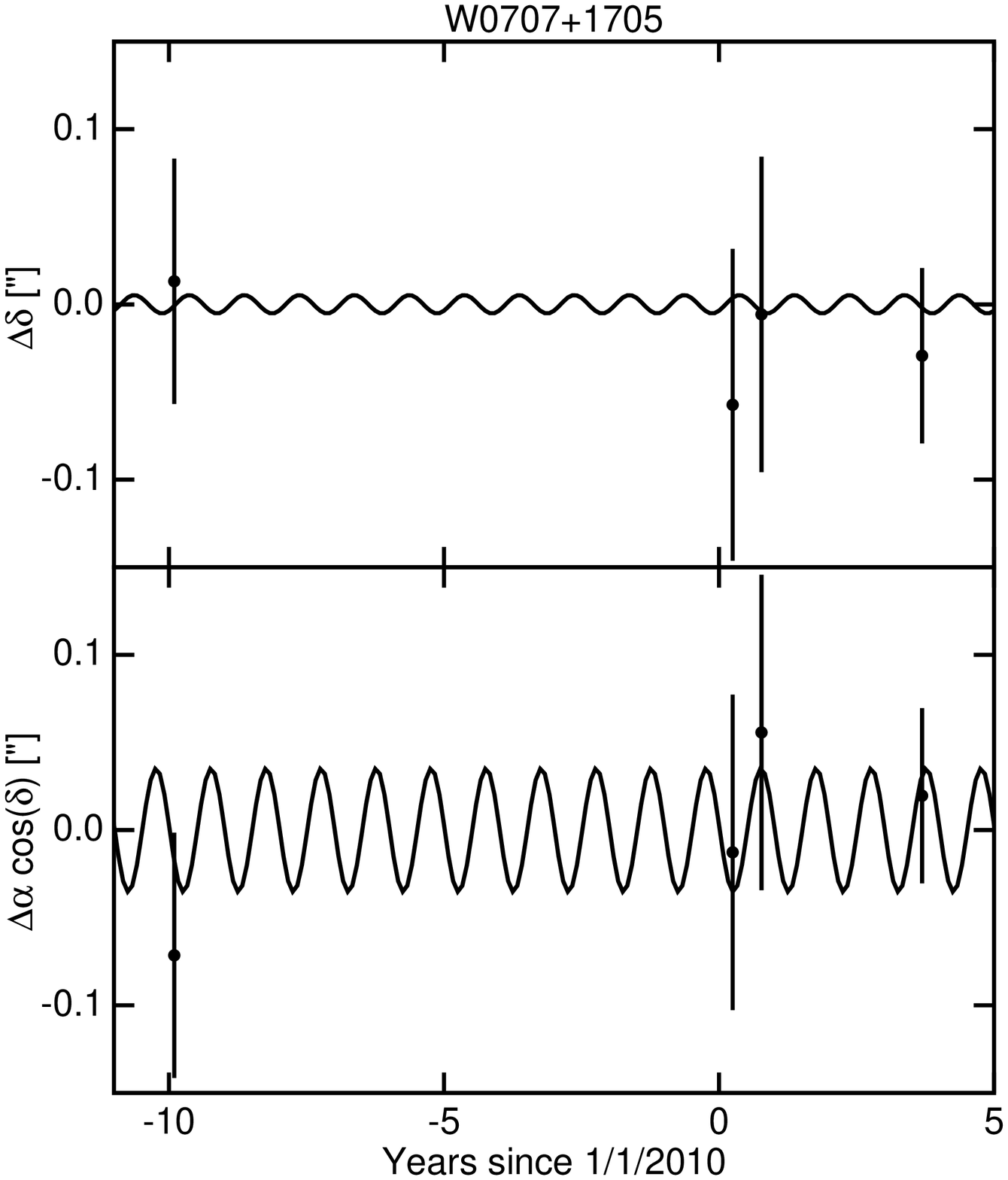}
\caption{Astrometric residuals after a proper motion only fit is removed.
Only the more precise data from 2MASS, WISE \& FanCam have been plotted
to reduce clutter.
\label{fig:W0707+1705-justpar}}
\end{figure}

\subsubsection{AllWISE}

The AllWISE catalog is formed by fitting the pixels in the individual
frames with a model containing 8 parameters which are
the position, motion and fluxes for a point source:
$(\alpha, \, \delta, \, \mu_\alpha = \cos\delta d\alpha/dt, \, \mu_\delta, \, 
F_{W1}, \, F_{W2}, \, F_{W3},
\, \& \, F_{W_4})$.  This model is convolved with the point spread
function of WISE, compared to the pixel data, and then adjusted
to get the best fit.  The position and motion
parameters, which have a non-linear effect on the
pixel values, are often hard to fit in a robust and stable way, so the
fitting procedure first finds the best fit with 
$\mu_\alpha = \mu_\delta  = 0$, which just
repeats the procedure used for the previous WISE catalogs, and
then uses this as a starting point 
for an iterative adjustment to find the best fitting motion.
In addition, the first $\mu = 0$ solution is used to generate the
source designation and when extracting fluxes from individual frames
that go into the multi-epoch photometry database (see \S\ref{sec:var}).

The AllWISE catalog entry for \obj\ has the designation
\AWP\ J070720.50+170532.7.  
The first position solution, for $\mu_\alpha = \mu_\delta  = 0$
gives  $\alpha = 106.8354461^\circ \pm 0.0446\asec$,
$\delta = 17.0924311^\circ\pm 0.0431\asec$. 
The magnitudes obtained in the fixed position solution are
W1234 $=12.646(24)$, $12.502(25)$, $ >11.701$ \& $ >8.719$.
The contamination and confusion flags are  all zero, indicating that
the detection is not affected by known artifacts.
This position is then used to initialize the simultaneous
solution for the position at epoch MJD=55400, the proper motion,
and the four WISE magnitudes, giving $\alpha =
106.8354766^\circ\pm0.0459\asec$, $\delta = 17.0924049^\circ\pm
0.0446\asec$, $\mu_\alpha = 1355\pm59$~mas/yr, $\mu_\delta =
-1256\pm62$~mas/yr, and W1234 $ = 12.660(23),\;12.512(25),\;12.430(530)\;
\& >8.702$.  Note that the W3 flux is essentially identical in both
solutions but one is slightly more than $2\sigma$ while the other
is slighty less and thus quoted as an upper limit.  The quoted
motion assumes that the parallax is zero, and a star with a non-zero
parallax will have a motion that differs from the true proper motion
by as much as $\pm 4\varpi$ since the parallactic movement in 6 months is $\pm
2\varpi$ and the time interval is one half year.

Note that the error on the motion of \obj\ is considerably smaller than the
error on the position.  This happens because the reference frame error terms
cancel out in the position difference.  For the optimal range of magnitudes,
$8 < W1 < 10$, with high SNR but not saturated, the motion errors can be as
small as 35 mas/yr, $1\sigma$, 1-axis.  Thus the smallest motions that can
reach $5\sigma$ are about 175 mas/yr.  But for fainter sources the noise terms
dominate the motion error, so for $W1 \approx 15$ the errors are $10 \times$ 
larger, giving a $5\sigma$ threshold of about 1750 mas/yr.  
The upper limit on motions that can be detected by AllWISE
is still being investigated and will be discussed further in
\citet{kirkpatrick/etal:2014}.
But Barnard's star and Kapteyn's star are recovered as moving sources, even
though they are quite saturated.

\subsubsection{Separated WISE epochs}

In order to combine the WISE astrometry with archival data from
earlier surveys, we use the position from the WISE Allsky catalog,
$\alpha = 106.8353609^\circ \pm0.0898\asec$, $\delta=   17.0925197^\circ\pm
0.0889\asec$ at a mean MJD of 55287.577.

Since \obj\ is bright enough to be easily detected on individual
frames, the mid-average of the positions found on 12 post-cryogenic
detections is used, giving $\alpha = 106.835568^\circ\pm0.09\asec$,
$\delta=  17.092344^\circ\pm0.09\asec$ at a mean MJD of 55479.133.

\subsubsection{Cross Identifications}

The AllWISE motion backtracked ten years to the 2MASS epoch predicts
a position NW of the WISE position.  Indeed,
an easily seen source in the 2MASS \citep{skrutskie/etal:2006} image
NW of the WISE position is not seen in the WISE image, while the
WISE source is not seen in the 2MASS image.   This source is 2MASS
J07071961+1705464 with $\alpha =   106.831722^\circ\pm0.07\asec$,
$\delta =  17.096224^\circ\pm0.07\asec$, and magnitudes JHK$_s =
13.473(25),\;13.045(29)\;\&\;12.923(17)$ on MJD = 51579.657.

\obj\ appears in the Carlsberg Meridian Catalogue 14 
\citep [CMC14]{evans/irwin/helmer:2002}. 
This observation also gives an AB magnitude of 
$15.811 \pm 0.17$ in the SDSS $r$ band, where the
quoted error is the external uncertainty for sources with $r_{AB} \approx 16$
\citep{evans/irwin/helmer:2002}.
The UCAC4 \citep{zacharias/etal:2013} contains \obj\ as entry
536-039034.  No proper motion is given.  The 579-642 nm magnitude
is $15.881 \pm 0.30$ on the Vega system.  These CCD-based surveys
give better photometry than photographic plates, but \obj\ is close
to the survey limit in both cases.

The USNO B1.0 catalog \citep{monet/etal:2003} contains several
entries that appear to refer to \obj.   Entries 1071-0154281,
1070-0147856 \& 1070-0147865 have positions and epochs that are
consistent with the movement of \obj, while the entry 1070-0147855
appears to be \obj\  in the second epoch blue plate associated with
an incorrect identification on the first epoch blue plate.  In
addition to the astrometric data, we take photometric values of B
$= 17.66,\; 17.70\; \& \; 18.33$, R $= 15.46$ and I $= 14.97$ from
these entries.  Some of these entries have large position errors
(999 max) but all are reasonably consistent with the astrometric solution.

Magnitudes from WISE, 2MASS and the USNO-B are on the Vega system.

In addition to the WISE,  2MASS \& USNO B astrometry, data were
also obtained by fitting parabolae to the data around the peak of
the source on FITS postage stamp images from the Digitized Sky
Survey maintained at the Space Telescope Science Institute.  These
values rely on the World Coordinate System in the FITS header and
the pixel positions reported by the SAOImage DS9 FITS viewer
\citep{joye/mandel:2003}.  For these values the error is taken to be
$1/3$ of a pixel in the FITS image, typically 333 mas.

\subsubsection{Fan Mountain Observatory/FanCam}

YJH\&K$_s$ photometry and astrometry of \obj\ were obtained on
5 Sep 2013
with FanCam, a HAWAII-1 based
near-infrared imager operating at the University of Virginia's Fan
Mountain 31-inch telescope \citep{kanneganti/etal:2009}. 
Results are given in Tables \ref{tab:astrometry} and \ref{tab:photometry}.

All of the astrometric data used in this paper are listed in Table
\ref{tab:astrometry}.  Figure \ref{fig:0707+1505-POSS12-W1} shows
images of \obj\ from 1952 to 2010.

\subsubsection{Ultraviolet}

\obj\ was not detected by GALEX \citep{martin/etal:1999} but its position
was observed during the Allsky Imaging Survey.  Examining
the expected position of \obj\ on the near UV FITS
image of the field shows it is at least 17 times fainter than
the nearby source at $\alpha = 106.837313^\circ,\;\delta= 17.099230^\circ$
which has a near UV AB magnitude of 18.32, so \obj\ is fainter than
21.4 or $< 10^{-5}$ Jy at 0.24 $\mu$m.

\subsubsection{Astrometric Fit}

A parallax and proper motion fit to the data in Table \ref{tab:astrometry} gives
$\mu_\alpha = 1228 \pm 3$ mas/yr, $\mu_\delta = -1307  \pm 1$ mas/yr, and
a parallax of $\varpi = 35 \pm 42$ mas.  The residuals give a $\chi^2$ per
degree of freedom $=8.72/19$, so the errors are reasonable if perhaps
conservative.
This best fit solution to a long time baseline
is consistent with the six month baseline AllWISE catalog motion
within $2\sigma$:
$\Delta\mu_\alpha = -127 \pm 59$ and $\Delta\mu_\delta = -51\pm62$
mas/yr.
Figure \ref{fig:W0707+1705-justpar} shows the residuals after a position
and proper motion fit is removed.

\subsubsection{Photometric Fit}

\begin{table}[tb]
\begin{center}
\caption{Photometric Observations of \obj.\label{tab:photometry}}
\begin{tabular}{lll}
\tableline\tableline
Filter & Magnitude & Instrument\\
\tableline
B & $17.9\pm0.25$ & USNO B\\
579-642 nm & $15.881 \pm 0.3$ & UCAC4\\
$r_{AB}$ & $15.811\pm0.17$ & CMC14\\
R & $15.46 \pm 0.33$  & USNO B\\
I & $14.97 \pm 0.33$  & USNO B\\
Y(MKO) & $13.90\pm0.01$ & FanCam \\
J &  $ 13.47 \pm 0.01 $ & FanCam \\
J &  $  13.473\pm 0.025$ & 2MASS \\
H & $ 13.08\pm0.01$ & FanCam \\
H & $ 13.045\pm0.029$ & 2MASS \\
K$_s$ & $ 12.91\pm0.01$ & FanCam \\
K$_s$ & $ 12.923\pm0.017$ & 2MASS \\
W1 & $12.660\pm0.023$ & \AWP \\
W2 & $12.512\pm0.025$ & \AWP \\
W3 & $12.430 \pm 0.530$ & \AWP \\
W4 & $ >8.702$ & \AWP \\
\tableline
\tablecomments{All magnitudes except
$r_{AB}$ are Vega magnitudes and\\ 
the JHK$_s$ bands use 2MASS filters.
The Y-band\\
calibration uses the \citet{hamuy/etal:2006} transformation\\
of Y-K$_s$ \vs\ J-K$_s$.}
\end{tabular}
\end{center}
\end{table}

\begin{figure}[tb]
\plotone{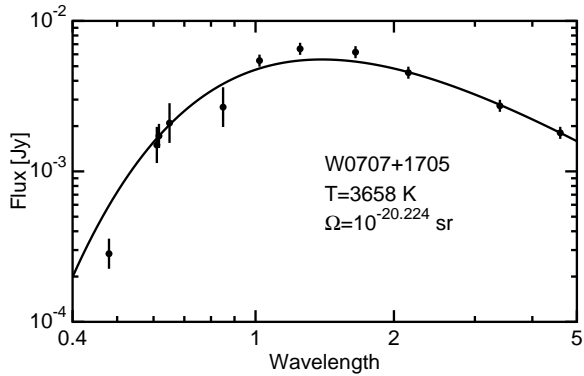}
\caption{Photometric points and a blackbody fit.
\label{fig:0707+1705-SED}}
\end{figure}

The magnitudes from WISE, 2MASS and the USNO B were fit
to a blackbody using a least sum of absolute value of error
criterion.   
\citet{monet/etal:2003} state that the photometric calibration of
the USNO B catalog is of marginal quality due to the lack
of an all-sky network of faint photometric calibration stars,
and that the standard deviation of the fit to the existing calibration
network is 0.25 mag.
Thus we used assumed uncertainties of 0.33 mag for the
USNO B except in the blue where there are three consistent
estimates so a 0.25 mag uncertainty was used. 
For WISE and 2MASS
the actual magnitude uncertainties were replaced by 0.1 mag to allow
for the crudeness of the blackbody model.  
The error for the CMC14 was rounded up to 0.2 mag for the same reason.
The photometric points
and the best fit model with $T = 3658$~K and a solid angle
of $10^{-20.224}$~sr are shown in
Figure \ref{fig:0707+1705-SED}.  This solid angle corresponds to
the Sun at a distance of 517 pc with an angular radius of 9 $\mu$as.

A more pertinent comparison is the moderately metal poor 
Barnard's star (GJ 699) which
has a measured radius of $0.1867 \pm 0.0012\;R_\odot$ 
\citep{boyajian/etal:2012}.  This would put \obj\ at a distance
of 97 pc.  A blackbody fit to the fluxes of Barnard's star,
similar to the one we did for \obj,
gives  a radius of 587 mas.  The distance of Barnard's star
is 1.8 pc, so if \obj\ were like Barnard's star its distance would
be 117 pc.

The near-infrared fluxes are clearly not suppressed by H$_2$
collision-induced absorption, as would be expected for a cool white
dwarf with a hydrogen atmosphere like LHS 3250 \citep{harris/etal:1999}.  
In fact, the slight excess over the
fit at J and H might indicate the presence of H$^-$ absorption.

\subsection{Variability\label{sec:var}}

\begin{figure}[tb]
\plotone{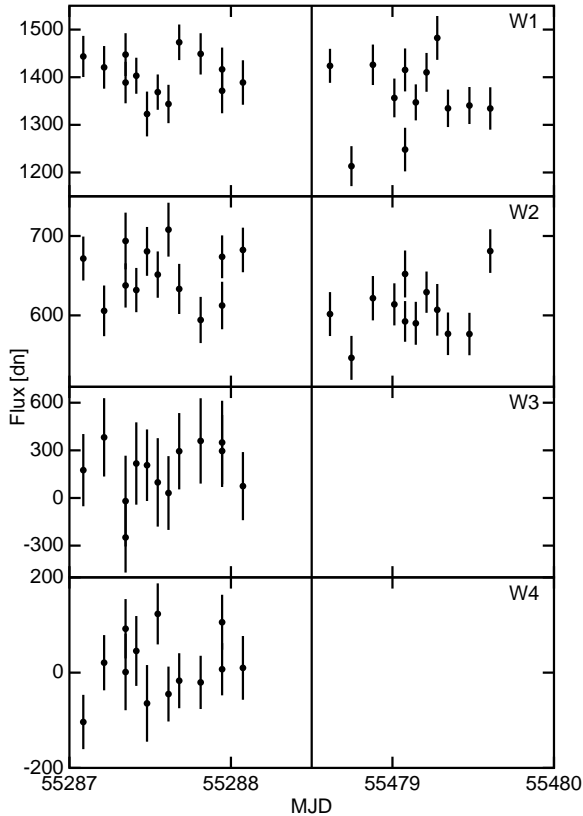}
\caption{Plot of multi-epoch photometry data for \obj.
In each band, the left and right panels cover the first
and seconds epochs of observation, separated by 6 months.
Fluxes are in data numbers.  1 dn corresponds to magnitudes
W1234 = 20.5, 19.5, 18 \& 13.
\label{fig:0707+1705_mep}}
\end{figure}

A comparison of the 2MASS and FanCAM magnitudes shows that
\obj\ has had a constant brightness over 13 years.  But a new product in
the AllWISE data release allows one to check for variability over hourly
or six month intervals.
In addition to checking for motion of sources when generating the catalog,
AllWISE is also producing a very large database of multi-epoch photometry.
This MEP database contains the flux estimates in each WISE band for each
frame that covers a source.  These individual frame flux estimates have been
generated in past releases, and used to set the variability flag, but not stored.
However, for AllWISE
the single frame values are kept in a searchable database
available at irsa.ipac.caltech.edu.
These flux estimates are forced photometry at the position of the source
assuming it is celestially fixed, not the moving position from the proper
motion fit.
Since the AllWISE catalog and reject table combined contain
more than $10^9$ sources,
and each source is observed on an average of 36 frames, the MEP database has
42,759,337,365 rows and thus should not be used for large area blind surveys
for variable objects.  But it is easy to collect the individual frame data for a source
or a list of sources.

The MEP database differs from the previously released
single frame detection list in one important
regard:  the single frame detections had to be $> 5\sigma$ in at least one band
in a single frame.  For sources that are just over $5\sigma$ in W1 in the coadd
of 30 frames, which includes a large fraction of catalogued WISE sources, the
individual frames will usually only yield upper limits.  The single
frame detection list will only have the few frames with upward noise fluctuations
leading to a biased view of the source flux.
Because most fluxes in the MEP will be insignificant, it is a good idea to use the
flux and flux sigma columns for analysis.  The fluxes are quoted in data numbers,
which are close to the actual quantization in the downloaded images, but the data numbers
have been scaled to correct for the different pixel and frame sensitivities, so 1 dn
always corresponds to a consistent flux.  The zeropoint magnitudes that correspond 
to a flux of 1 dn are W1234 = 20.5, 19.5, 18 \& 13, the same as the zeropoints for the 
allsky release given in Table 1 of \S 2.3.f of the Explanatory Supplement.

Figure \ref{fig:0707+1705_mep} shows the flux values and uncertainties in the
4 WISE bands for the 25 frames covering \obj.  Fitting a constant flux to each band
gives $1383.10 \pm  8.38$~dn with a $\chi^2$ of 54.8 for 24 degrees of freedom
in W1, $627.93 \pm 5.72$~dn with $\chi^2/\mathrm{dof} = 49.9/24$ in W2,
$162.7 \pm 67.3$~dn with $\chi^2/\mathrm{dof} = 7.0/12$ in W3, and
$11.24 \pm 17.29$~dn with $\chi^2/\mathrm{dof} = 14.2/12$ in W4.
These uncertainties are statistical only and
do not include systematic terms such as the uncertainty in
the zeropoint.  The W3 value corresponds to W3 $= 12.47$ with an
SNR of 2.42:1, providing no evidence for a warm dust disk around \obj.

Moderate improvements in the $\chi^2$ for W1 and W2 can be made by
fitting the epochs separately.  $\Delta\chi^2 = 6.4$ in W1 with the
later epoch $\Delta m = 0.033$ mag fainter.  In W2, $\Delta\chi^2$ is 14.9
with the second epoch 0.076 mag fainter.  Given all the changes in
the detector operating conditions as the telescope warmed up, these results
are only weak evidence for  a small amount of long term variability
over a 6 month time span.  But the comparison of 2MASS and FanCam over
a 13 year span indicates that \obj\ is probably not varying.

\subsection{Spectroscopy}\label{sec:spectroscopy}

\begin{figure}[tb]
\plotone{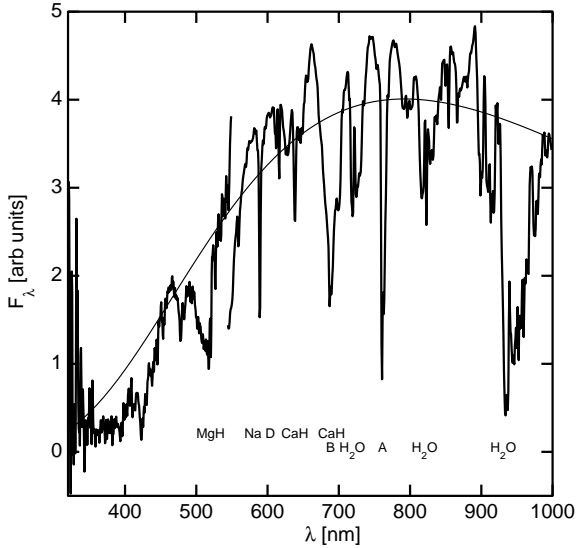}
\caption{Palomar Double Spectrograph spectrum of \obj\ is shown as the
heavy dark line.  The discontinuity near
550 nm is due to the dichroic separating the red and blue sides of the
spectrograph.  
This spectrum has been smoothed to the resolution of the plot with
a 1.5 nm FWHM smoothing kernel.
The lower line of spectral identifications is due to
telluric absorption bands.  
An arbitrarily scaled 3680~K blackbody
gives the thin solid curve.
\label{fig:0707+1705-spectrum}}
\end{figure}

\begin{figure}[tb]
\plotone{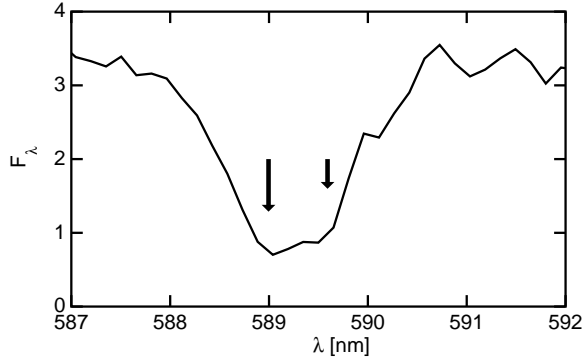}
\caption{Full resolution blowup of the Na D line region of the \obj\ spectrum.
The arrows indicate the rest wavelengths of the doublet.
\label{fig:0707+1705-NaD-spectrum}}
\end{figure}

\begin{figure}[tb]
\plotone{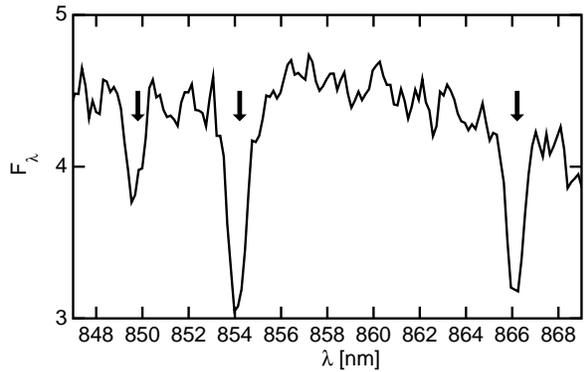}
\caption{Expanded view of the Ca II IR triplet part of the spectrum.
\label{fig:0707+1705-Ca+IR-spectrum}}
\end{figure}

\begin{figure}[tb]
\plotone{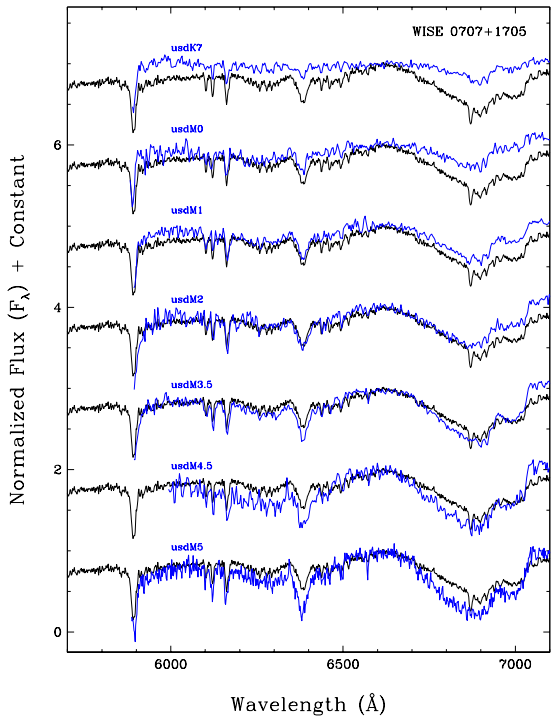}
\caption{Spectrum of \obj\ compared to spectral standards
from \cite{lepine/rich/shara:2007}.  All spectra are normalized
at 660 nm.
\label{fig:WISE0707+1705_FigA}}
\end{figure}

\begin{figure}[tb]
\plotone{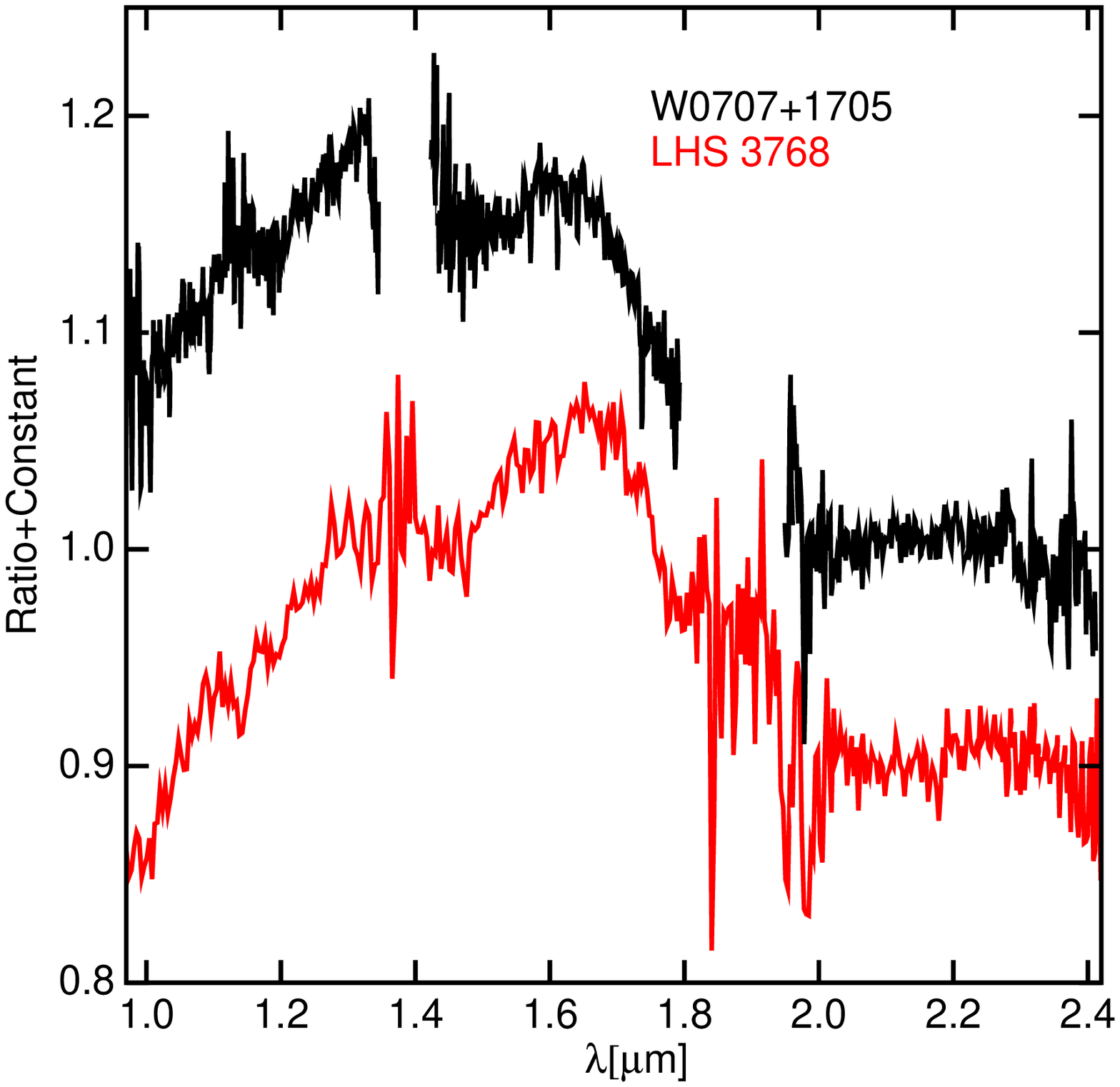}
\caption{The Triple-Spec spectrum divided by the blackbody fit,
compared to a SpeX spectrum of LHS 3768.  The spectra are
normalized at 2.2 $\mu$m, and LHS 3768 is displaced downward
by 0.1 for clarity.
\label{fig:0707+1705-Tspec-ratio}}
\end{figure}

We obtained an optical spectrum on UT 2013 September 3 with the Double
Spectrograph at Palomar Mountain Observatory through partial
cloud coverage during morning twilight.  This spectrum is plotted
in Figure \ref{fig:0707+1705-spectrum}.
No calibration star could be observed so the sensitivity calibration is based
on calibrations over the previous month, and atmospheric features
have not been divided out.
The 3680 K blackbody fit to the photometric data is a reasonable match
to the trend of the spectrum, but the flux scaling is arbitrary.

The sodium D doublet of \obj\ is deep and broad with an equivalent
width of $\approx 1.2$ nm and a FWHM for the merged lines of $\approx 1.4$ nm.
It shows a radial velocity $< 100$ km/s,
as seen in Figure \ref{fig:0707+1705-NaD-spectrum}.
The Ca II infrared triplet indicates $v_{rad} = -45 \pm 18$~/km/sec
with respect to the Earth, or -21~km/sec with respect to the Sun.  
Figure \ref{fig:0707+1705-Ca+IR-spectrum}
gives an expanded view of the Ca II infrared triplet.

The TiO band at 715 nm is much weaker than the 690 nm CaH band, and
the telluric water band at 718.5 nm may provide all the structure seen
in this region of the spectrum.  Weak
TiO to CaH is an indicator of an extreme or ultrasubdwarf,
according to \citet{lepine/rich/shara:2007}.   Figure
\ref {fig:WISE0707+1705_FigA} compares the spectrum of \obj\ to the spectral
standards in \citet{lepine/rich/shara:2007}.  
In this region \obj\ resembles an usdM2 or usdM3.5 star.
The usdM3.5 spectral standard is LHS 325, while the usdM2 standard
is LHS 1691.  Their resemblance to \obj\ includes similar spectral energy 
distributions: blackbody fits to the USNO B, 2MASS and WISE data give
3499 K for LHS 325 and 3184 for LHS 1691, which are similar to the fit for \obj.

The sequence from ordinary red dwarf (dM), through subdwarf (sdM), extreme
subdwarf (esdM) to ultra subdwarf (usdM) is one of weakening metal oxide
bands relative to metal hydride bands and atomic lines.  Thus this is a sequence 
a decreasing metallicity, since to make a TiO molecule in a very low metallicity
star requires two very rare atoms.

The near infrared CN and CO bands
are not seen in a spectrum obtained on UT 2013 September 15 with the
Triple-Spec instrument on the Apache Point Observatory
3.5 m telescope.   The flux in the spectrum has been normalized
using the 2MASS photometry.  Figure \ref{fig:0707+1705-Tspec-ratio}
shows the spectrum divided by the blackbody fit in 
Figure \ref{fig:0707+1705-SED}.  This is compared to a NASA IRTF
SpeX prism spectrum of LHS 3768, which is tentatively classified
as usdM3 based on the spectrum in
\citet{kirkpatrick/henry/simons:1995}.  These spectra are very similar,
and neither shows any strong molecular bands.
The CN band at 1.1 $\mu$m in the spectrum of the carbon dwarf G77-61 shown by
\citet{joyce:1998} is at least ten times stronger than in \obj.
The iron abundance of G77-61 is very low, $10^{-4}$ of solar, but both carbon
and nitrogen are enhanced relative to iron by a factor of 400 \citep{plez/cohen:2005}.
The weakness of any bands from carbon containing molecules in \obj\ suggests
that carbon is depleted along with the metals in this source.

\section{Discussion}

The spectral standards LHS 325 and LHS 1691 do not have published
precise parallaxes so it is difficult to estimate the distance and
absolute magnitude of \obj.  But
\citet{bochanski/etal:2013} gives estimated absolute magnitudes for red dwarfs
and subdwarfs seen by the SDSS \citep{york/etal:2000}.  This paper defines a
standard color $(g-r)_{dM}$ for a red dwarf as a function of the $(r-z)$ color,
and then uses the color excess $\delta(g-r) = (g-r)-(g-r)_{dM}$ as an indicator
of subdwarfism.
\citet{bochanski/etal:2013} then gives fits for the absolutes magnitudes M$_{r,i,z}$
as a function of $(r-z)$ and $\delta(g-r)$.
Unfortunately we do not have SDSS $griz$ photometry on \obj, but we do have
SDSS data on LHS 325, with
{\it ugriz} $=21.59,\;17.96,\;16.08,\; 15.26\;\&\; 14.81$ in DR10 
\citep{ahn/etal:2013}.  These data
imply $\delta(g-r) = 0.49$, appropriate for an ultra-subdwarf, and an
absolute magnitude of $M_r = 12.883 \pm 0.41$ based on the
polynomial fits in \citet{bochanski/etal:2013}.
If \obj\ has the same absolute magnitude, then
the apparent $r$ magnitude (AB system) of  $\approx 15.811 \pm 0.17$ from the
CMC14 data implies a distance of $39 \pm 9$ pc and a tangential velocity
of $v_{tan} = 330$~km/sec.  This seems unlikely, given the small radial velocity,
but it is not ruled out.  The parallax determination is consistent with $d = 39$ pc, but
has such a large uncertainty that it does not improve our knowledge of the distance,
except to impose a lower limit, $d > 6$~pc.

Note that the proper motion of \obj\ is primarily an
increasing galactic longitude ($\cos(b)dl/dt = 1705$~mas/yr,
$db/dt = 555$~mas/yr), 
as would be expected for a source not rotating with
the Galaxy, since its location is near the anti-center [$(l,b) =  (199.29^\circ,11.13^\circ)$].
The radial velocity of $-21$~km/sec with respect to the Sun gives
UVW components $(19, 7, -4)$~km/sec, 
and the proper motion gives $(3.15d,-7.46d,+2.58d)$~/km/sec with $d$ in pc,
while the Sun's velocity with respect to the Local Standard of Rest (LSR)
is $(9,12,7)$~km/sec \citep{binney/tremaine:1987}, so the total velocity with respect 
to the LSR is $(28+3.15d, 19-7.46d, 3+2.58d)$~km/sec.

If \obj\ is a member of a non-rotating singular isothermal
sphere halo population, 
with $\sigma = v_c/\sqrt{2}$ where $v_c \approx 220$~km/sec is the
local circular velocity, then the velocity distribution function is
\bea
f(U,V,W) & \propto & \exp[-(U^2+W^2-(V+v_c)^2)/v_c^2] \nonumber \\
              & \propto & \exp[-(1.2-0.070d+0.00147d^2)].
              \nonumber \\
              & &
\eea
The argument of the exponential is only $-0.715$ for $d = 39$~pc.
If \obj\ is 1.1 mag less luminous than we have assumed, the
argument of the exponential reaches its maximum value of
$-0.371$ at $d = 24$~pc.  This is equivalent to a Gaussian in
$d$ with a mean of 24 pc and a standard deviation of 18 pc.

But a singular isothermal sphere velocity distribution is an
extreme assumption with an infinite escape velocity.  
\citet{bochanski/etal:2013} also derive
an ellipsoidal velocity distribution for usdM stars.  For this distribution, 
the peak is at $d = 15$ pc and $\sigma$ is 15 pc.  This
case is more in favor of \obj\ being 2 magnitudes more sub-luminous
than an ultra-subdwarf star, but even the $d = 39$ pc case with 
$v_{tan} = 330$ km/sec is only $1.7\sigma$ out in the tail of this
distribution.

For $d = 39$ pc \obj\ is in a retrograde orbit around the Milky Way, and its speed
with respect to the center of mass of the galaxy is no more than 
the local rotational speed.  Thus there is no indication that \obj\ is
escaping from the Milky Way even if $v_{tan} \approx 330$~km/sec.

An accurate parallax would certainly be useful to 
distinguish between these scenarios for the luminosity and
distance of \obj.
Improved optical photometry and spectroscopy would help to put \obj\
into the context of the many faint red stars seen by large surveys
such as the SDSS.

\section{Conclusions}

\obj\ shows the power of the upcoming AllWISE data release to reveal
odd denizens of the Solar neighborhood.  With an extremely short
interval between epochs, the source association problems that have
plagued previous searches for large proper motion objects are
eliminated.  Some examples based on discoveries within
the last decade are:
Teegarden's star \citep{teegarden/etal:2003} is recovered with
AllWISE proper motion values of $\mu_\alpha = 3471 \pm 56$
\& $\mu_\delta = -3666 \pm 25$ mas/yr;
WD 1339-340 \citep{lepine/rich/shara:2005} is recovered with
 $\mu_\alpha = -1916 \pm 78$
\& $\mu_\delta = 999 \pm 84$ mas/yr; and
Luhman's star \citep{luhman:2013} is recovered with
 $\mu_\alpha = -2646 \pm 26$
\& $\mu_\delta = 408 \pm 25$ mas/yr.
These stars are all easily visible on optical sky survey
plates but images at different epochs were not correctly associated.
Previously known higher proper motion stars like Barnard's star
and Kapteyn's star are also easily recovered.
See \S II.6.d of the AllWISE Explanatory Supplement for more information 
on the AllWISE motion limits.

Based on sampling in test regions, we estimate that
2000-3000 real new proper motion detections will be found in the whole
sky \citep{kirkpatrick/etal:2014}.  
In addition to proper motions, AllWISE will provide a deeper
survey of the sky at 3.4 \& 4.6 \um, and new tools to study the
variability of sources all over the sky.

\acknowledgments

We thank the referee, Ralf-Dieter Scholz, for a very fast and useful report
pointing out the UCAC and CMC observations.

This publication makes use of data products from the Wide-field
Infrared Survey Explorer, which is a joint project of the University
of California, Los Angeles, and the Jet Propulsion Laboratory/California
Institute of Technology, funded by the National Aeronautics and
Space Administration.

The WISE, 2MASS and USNO B data were all provided by the Infrared
Science Archive at Caltech.

The Digitized Sky Survey images were provided by the Space Telescope
Science Institute.

Support for MF was provided by NASA through Hubble Fellowship grant 
HF-51305.01-A

This research has benefitted from the SpeX Prism Spectral Libraries maintained by Adam Burgasser at http://www.browndwarfs.org/spexprism

{\it Facilities:} \facility{WISE}, 
\facility{Fan Mountain/FanCam}, 
\facility{Palomar/DoubleSpec},
\facility{ARC/TSpec}


\end{document}